\documentclass[english,aps,superscriptaddress,showkeys,showpacs,prepri,twocolumn]{revtex4}
\usepackage[T1]{fontenc}
\pdfoutput=1
\usepackage[letterpaper]{geometry}
\usepackage{units}
\usepackage{bbding}
\usepackage{amsmath}
\usepackage{amssymb}
\usepackage{graphicx}
\usepackage{esint}
\usepackage[colorinlistoftodos,prependcaption,textsize=tiny]{todonotes}

\usepackage[utf8]{inputenc}
\usepackage{lmodern} % load a font with all the characters

\makeatletter

\@ifundefined{textcolor}{}
{%
 \definecolor{BLACK}{gray}{0}
 \definecolor{WHITE}{gray}{1}
 \definecolor{RED}{rgb}{1,0,0}
 \definecolor{GREEN}{rgb}{0,1,0}
 \definecolor{BLUE}{rgb}{0,0,1}
 \definecolor{CYAN}{cmyk}{1,0,0,0}
 \definecolor{MAGENTA}{cmyk}{0,1,0,0}
 \definecolor{YELLOW}{cmyk}{0,0,1,0}
}

\usepackage{doi}
\usepackage{hyperref}

\makeatother

\usepackage{babel}
\begin{document}

\title{
The Impact of Collisionality, FLR and Parallel Closure Effects on Instabilities
in the Tomakak Pedestal:   Numerical Studies with the NIMROD code
}
%Edge NIMROD studies and verification:\\ideal-MHD modification by collisionality, \\FLR and parallel-closure effects.}

\author{J. R. King}
\affiliation{Tech-X Corporation, 5621 Arapahoe Ave. Boulder, CO 80303, USA}

\author{A. Y. Pankin}
\affiliation{Tech-X Corporation, 5621 Arapahoe Ave. Boulder, CO 80303, USA}

\author{S. E. Kruger}
\affiliation{Tech-X Corporation, 5621 Arapahoe Ave. Boulder, CO 80303, USA}

\author{P. B. Snyder}
\affiliation{General Atomics, PO Box 85608, San Diego, CA 92186–5608, USA}

\begin{abstract}
The extended-MHD NIMROD code [C.R.~Sovinec and J.R.~King, J.~Comput.~Phys.~{\bf
229}, 5803 (2010)] is verified against the ideal-MHD ELITE code [H.R.~Wilson
\textit{et al.}~Phys.~Plasmas {\bf 9}, 1277 (2002)] on a diverted tokamak
discharge. When the NIMROD model complexity is increased incrementally,
resistive and first-order finite-Larmour radius effects are destabilizing and
stabilizing, respectively. The full result is compared to local analytic
calculations which are found to overpredict both the resistive destabilization
and drift stabilization in comparison to the NIMROD computations.
Published version: Phys. Plasmas 23, 062123 (2016) [\url{http://dx.doi.org/10.1063/1.4954302}]
\end{abstract}


\maketitle

%--------------------------------------------------

%These are definitions of new commands

%GENERAL PURPOSE
 % Define how vectors are annotated
 % \newcommand{\vect}[1]{ \vec{#1}}
  \newcommand{\vect}[1]{ \mathbf{#1}}
 %These are used for defining variables
  \newcommand{\defn}{ \equiv}

 %Left and Right Delimiters
  \newcommand{\lp}{\left(}
  \newcommand{\rp}{\right)}
  \newcommand{\lb}{\left[}
  \newcommand{\rb}{\right]}
  \newcommand{\la}{\left<}
  \newcommand{\ra}{\right>}

 % Miscellaneous
  \newcommand{\vf}{ \vect{f}}

 % MHD Variables
  \newcommand{\vx}{\vect{x}}
  \newcommand{\vq}{\vect{q}}
  \newcommand{\vB}{\vect{B}}
  \newcommand{\vJ}{\vect{J}}
  \newcommand{\vA}{\vect{A}}
  \newcommand{\vE}{\vect{E}}
  \newcommand{\vV}{\vect{V}}
  \newcommand{\vF}{ \vect{F} }	
  \newcommand{\vU}{ \vect{U} }	
  \newcommand{\ddp}{\grad \cdot \Pi}
  \newcommand{\specheat}{\gamma_h}

 % Grad, Curl, Divergence and other vector operators
  \newcommand{\grad}{\vect{\nabla}}
  \newcommand{\curl}[1]{\grad \times #1 }
  \newcommand{\dive}[1]{\grad \cdot #1 }
  \newcommand{\vdg}{\left(\vV \cdot \grad \right)}
  \newcommand{\bdg}{\left(\vB \cdot \grad \right)}
  \newcommand{\divV}{\grad \cdot \vV_1}
  \newcommand{\divVp}{\left( \grad \cdot \vV \right)}

 % Partial derivatives 
  \newcommand{\dt}[1]{\frac{\partial #1}{\partial t}}
  \newcommand{\Dt}[1]{\frac{d #1}{dt}}
  \newcommand{\dpsi}[1]{\frac{\partial #1}{\partial \psi}}
  \newcommand{\dpsisq}[1]{\frac{\partial^2 #1}{\partial \psi^2}}
  
% Vector Coordinate systems
  \newcommand{\jac}{{\mathcal{J}}}
  \newcommand{\jaci}{{\mathcal{J}}^{-1}}
  \newcommand{\Pp}{ P^\prime }				
  \newcommand{\Vp}{V^\prime}
  \newcommand{\Vpp}{V^{\prime\prime}}
  \newcommand{\Vpo}{ \frac{V^\prime}{4 \pi^2}}
  \newcommand{\norm}{ P^\prime }		% Normalization
  \newcommand{\RR}{ \psi }			% Radial coordinate
  \newcommand{\vR}{ \grad \RR }		% Radial coordinate
  \newcommand{\C}{ C }				% B x Grad \RR
  \newcommand{\vC}{ \vect{\C} }		% B x Grad \RR
  \newcommand{\vK}{ \vect{K} }			% J x Grad \RR
  \newcommand{\vRsq}{ \mid \grad \RR \mid^2 }
  \newcommand{\vCsq}{ \C^2 }
  \newcommand{\vKsq}{ K^2 }
  \newcommand{\vBsq}{ B^2 }
  \newcommand{\vrr}{\frac{ \vR}{\vRsq} }
  \newcommand{\vbb}{\frac{ \vB}{B^2} }
  \newcommand{\vcc}{\frac{ \vC}{\vCsq} }
  \newcommand{\vjj}{\frac{ \vJ}{J^2} }
  \newcommand{\vkk}{\frac{ \vK}{\vKsq} }

% rho, theta, zeta, (alpha, u) coordinate system
  \newcommand{\R}{ \psi }
  \newcommand{\T}{ \Theta }
  \newcommand{\Z}{ \zeta }
  \newcommand{\A}{ \alpha }
  \newcommand{\U}{ u }
  \newcommand{\ve}{ \vect{e} }
  \newcommand{\vur}{ \vect{e}^\rho }
  \newcommand{\vut}{ \vect{e}^\Theta }
  \newcommand{\vuz}{ \vect{e}^\zeta }
  \newcommand{\vlr}{ \vect{e}_\rho }
  \newcommand{\vlt}{ \vect{e}_\Theta }
  \newcommand{\vlz}{ \vect{e}_\zeta }
  \newcommand{\gr}{ \grad \R }
  \newcommand{\gt}{ \grad \Theta }
  \newcommand{\gz}{ \grad \zeta }
  \newcommand{\ga}{ \grad \alpha }
  \newcommand{\gu}{ \grad \U }
  \newcommand{\dr}[1]{ \frac{\partial #1}{\partial \R} }
  \newcommand{\dT}[1]{\frac{\partial #1}{\partial \Theta}}
  \newcommand{\dz}[1]{\frac{\partial #1}{\partial \zeta}}
  \newcommand{\dU}[1]{\frac{\partial #1}{\partial \U}}
  \newcommand{\drs}[1]{ \frac{\partial^2 #1}{\partial \R^2} }
  \newcommand{\dTs}[1]{\frac{\partial^2 #1}{\partial \Theta^2}}
  \newcommand{\drt}[1]{\frac{\partial^2 #1}{\partial \R \partial \Theta}}
  \newcommand{\dzs}[1]{\frac{\partial^2 #1}{\partial \zeta^2}}
  \newcommand{\grr}{ g^{\R \R} }
  \newcommand{\grt}{ g^{\R \Theta} }
  \newcommand{\grz}{ g^{\R \zeta} }
  \newcommand{\gtz}{ g^{\Theta \zeta} }
  \newcommand{\gtt}{ g^{\Theta \Theta} }
  \newcommand{\gzz}{ g^{\zeta \zeta} } 
  \newcommand{\ri}{ \frac{1}{R^2} }
  \newcommand{\fr}{ \lp \R \rp}
  \newcommand{\frt}{ \lp \R, \T \rp}
  \newcommand{\frtz}{ \lp \R,\T,\Z \rp}

  \newcommand{\fluxav}[1]{\la #1 \ra}
  \newcommand{\thetaav}[1]{\la #1 \ra_\T}

\newcommand{\cramplist}{
        \setlength{\itemsep}{0in}
        \setlength{\partopsep}{0in}
        \setlength{\topsep}{0in}}
\newcommand{\cramp}{\setlength{\parskip}{.5\parskip}}
\newcommand{\zapspace}{\topsep=0pt\partopsep=0pt\itemsep=0pt\parskip=0pt}

   % Useful abbreviations
\section{Introduction}
\label{sec:introduction}

Understanding the complex physics of instabilities in the tokamak pedestal
region is essential for ensuring that tokamaks achieve their highest performance.
Computation plays a critical role in furthering the knowledge of edge dynamics,
including advances in the understanding of type-I edge-localized modes
(ELMs)~\cite{leonard06}.  Nonlinear computation of ELM dynamics remains
challenging, but is becoming tractable with modern computing
resources~\cite{pankin07,sugiyama10,xu10,xu13,xi13,huijsmans13,pamela13}.  ELM
% Note that "linear" is a mathematical concept so applies to modeling
% rather than dynamics; i.e., phenomenologically you cannot measure a
% "linear phase" (you can infer that it is in the linear phase, but it's an inference).
modeling can be roughly broken into three regimes: the early linear regime, the
nonlinear regime, and an energy loss regime.  Two important aspects of the
linear stage are the stability boundary within an operational space;
e.g., pedestal current and pressure gradient; and the linear growth rate for each
toroidal-mode number; i.e.,  the growth-rate or mode spectrum.
The dominant instability drive that sets the linear-stability boundary is from the
peeling-ballooning mode (PBM)~\cite{connor98,Snyder02,snyder04} based on
the ideal-MHD model.
The linear spectrum is critical to the next phase, the nonlinear regime as it
seeds the dynamics.  Non-ideal effects can greatly impact the growth-rate
spectrum and are thus important in determining the nonlinear ELM dynamics.  In
the final regime, high-fidelity simulation of the energy loss requires modeling
of the fast conductive and convective transport, and accurate coupling to
neutrals and the wall.  State-of-the-art computations are beginning to push the
boundary of current capabilities to modeling that captures accurate dynamics
with these additional effects that are relevant during energy
loss~\cite{huijsmans13,pamela13,gui14}.  

As the subsequent nonlinear evolution and energy loss in simulations is greatly
impacted by the amplitudes of initial linear perturbations; a large source of
uncertainty arises from the linear phase of ELM evolution.  Given the
difficulty (or impossibility) of measuring the small perturbations that seed an
ELM, a reasonable model is to seed it from unstable PBMs that grow from
small-wave perturbations.  Although the stability boundary is well described by
ideal MHD, details of the mode spectrum depend upon specific non-ideal effects.
In particular, resistive effects may destabilize and first-order
finite-Larmour-radius (FLR) effects may stabilize the modes. These latter effects,
part of which is drift stabilization, require two-fluid modeling with advanced
FLR closures. It is important to verify that codes accurately capture these
effects within regimes of interest. In this paper, we study these effects on
the linear modes with the extended-MHD NIMROD code \cite{Sovinec04,Sovinec10}.

Type-I ELM stability boundaries are dominated by drives described by ideal MHD
as the ideal-MHD operator is stiff; i.e., near the ideal-MHD stability boundary
small changes in pressure and current-density gradients lead to large changes
in the growth rate due to the ideal-MHD terms becoming large. Thus only small
profile changes are required to cross the boundary, and non-ideal modifications
to the growth rate become secondary when describing the boundary location
within operational phase space. This ideal-MHD stiffness is a general property
that impacts the behavior of a wide class of plasma-instability phenomena.  For
example, in addition to ELM stability, it also describes the dynamics of
disruptions caused by core-mode activity \cite{callen99,kruger05}.  However,
type-I ELMs are more complicated than core-mode disruptions given the broad
toroidal-mode spectrum and associated nonlinear coupling. Similar to studies of
core-mode disruptions, comprehensive understanding of ELM onset requires
evolving the equilibrium on the transport time scale through the instability
boundary with a verified code that includes non-ideal terms to calculate an
accurate initial linear-mode spectrum.

MHD codes that study ELM instabilities can be placed into two broad
categories: 1) extended-MHD codes (e.g. NIMROD and
M3D-C1~\cite{jardin2007high}) and 2) reduced-MHD-based codes (e.g.
BOUT++~\cite{dudson2009bout} and JOREK~\cite{huysmans2009non}).  The
definition of extended MHD used in this work includes the full ideal-MHD
force operator (see Ref.~\cite{Schnack:2006kx} and references therein).
Using the full operator retains the fast compressional-Alfvén waves that
enforce force balance and thus are critical to extended-MHD transport
calculations with sources and sinks (e.g. \cite{Jenkins10}).  Relative to
reduced modeling, the extended-MHD approach also has advantages in
avoiding low-$\beta$ approximations and a straight-forward
implementation of the FLR closures as discussed in
Section~\ref{sec:xMHD}.  However, the inclusion of the full force
operator comes at a cost of increased numerical difficulties when
solving the equations associated with the fast
waves~\cite{gruber2012finite, degtyarev1986methods,
bondeson1991tunable,lutjens1996class, Sovinec15}.   Verification for all
regimes is important, and previous work~\cite{Burke10} verified the ideal-MHD
terms for a non-diverted, peeling-ballooning unstable test case.  As one
expects that transport calculations across the ELM stability boundary coupled
with an instability model that captures FLR drift stabilization significantly
impacts dynamics, emphasis on verifying linear physics of extended-MHD codes
thus naturally follows.

% One of the methods that NIMROD has developed for resolving these
% difficulties is to have the equilibrium terms separated from the
% dynamical terms.  The implications of this separation is discussed in
% the next section. 
% JRK - add this to separate EQ discussion. 

In this work, a verification of the full extended-MHD, initial-value NIMROD
code with the ideal-MHD ELITE code~\cite{wilson02,Snyder02,snyder04} is
performed using the Meudas-1-case comparison~\cite{Snyder09}. This is a
diverted tokamak case that is peeling-ballooning unstable. As we later
discuss, the inclusion of the x-point significantly modifies the mode
structure; making the comparison in full magnetic geometry an important
additional verification.  To assess the impact of the two-fluid, FLR effects,
we compare our results with drift stabilization to a calculation of drift
stabilization on the resistive PBM from Ref.~\cite{Hastie03}.  Although the
approximations made during the analysis of first-order FLR PBMs in
Ref.~\cite{Hastie03} preclude quantitative comparison, we make a qualitative
comparison of our first-order FLR computations to analytics.  

The paper proceeds as follows: In Sec.~\ref{sec:benchmark} we summarize the
benchmark and give an overview of our methodology.  The exhaustive details of
the methods used in this comparison, in particular the mechanism whereby a
vacuum region is effectively included through extended-MHD modeling, are
previously available in Refs.~\cite{Burke10} and~\cite{Ferraro10} and thus our
description of this aspect is brief.  In Sec.~\ref{sec:xMHD} the model
complexity is gradually increased from the ideal-like model through a
resistive-MHD model with experimentally relevant profiles to a full
extended-MHD model which includes both parallel-closure and FLR-drift effects.
The resulting impact on the growth-rate spectrum in analyzed. Comparisons
of the computed drift-stabilized growth rates are made to analytic treatments
of the PBM in Sec.~\ref{sec:analyticComp}.
In Sec.~\ref{sec:densityScan} the density and temperature are varied at fixed
plasma $\beta$ (thus the ion gyroradius, $\rho_i$, is modified) in order to
ascertain how the FLR effects are impacted.  In this study, the bootstrap
current is fixed (the zeroth-order effect from variations of density modify the
bootstrap current) and the impact of FLR drift effects are isolated.
Sec.~\ref{sec:conclusions} summarizes this work and places it within the
broader context of modeling of ELM relaxation events in the tokamak edge with
transport effects.

\section{Verification benchmark}
\label{sec:benchmark}

We begin with a study of a high resolution, lower-single-null, JT-60U-like
equilibrium (`Meudas-1'), which was originally employed in a benchmark of the
MARG2D and ELITE codes \cite{Aiba07}, including a
close approach to the X-point \cite{Snyder09}.
This extends previous benchmarks~\cite{Burke10} of ELITE and NIMROD as it
includes diverted magnetic topology and a higher edge safety factor
($q_{95}=6.74$, the safety factor at 95\% of the normalized poloidal flux) that
leads to increased resolution requirements. An ideal-MHD limit is achieved in
NIMROD by using flat density and resistivity profiles inside the last closed
flux surface (LCFS) with small resistivity, $S=10^8$ where $S$ is the Lundquist
number  ($S=\tau_R/\tau_A$), $\tau_A$ is the Alfvén time ($\tau_A=R_o/v_A$),
$v_A$ is the Alfvén velocity ($B/\sqrt{m_i n_i \mu_0}$), $\tau_R$ is the
resistive diffusion time ($\tau_R=R_o^2 \mu_0/\eta$), $R_o=2.936 m$ is the
radius of the magnetic axis, $\eta$ is the electrical resistivity, $\mu_0$ is
the permeability of free space, $m_\alpha$ is a species mass (the $\alpha$
subscript denotes ions or electrons in this work), and $n_\alpha$ is a species
density. The deuteron mass ($m_i = 3.34\times 10^{-27} kg$) is used. In order to
reproduce the vacuum response model outside the LCFS that is used by ELITE, a
low density ($0.01$ of the core density) and high resistivity ($10^7$ times the
core resistivity) is prescribed beyond the LCFS (more details on these
approximations are in Ref.~\cite{Burke10}).  

\begin{figure}
  \centering
  \includegraphics[width=8cm]{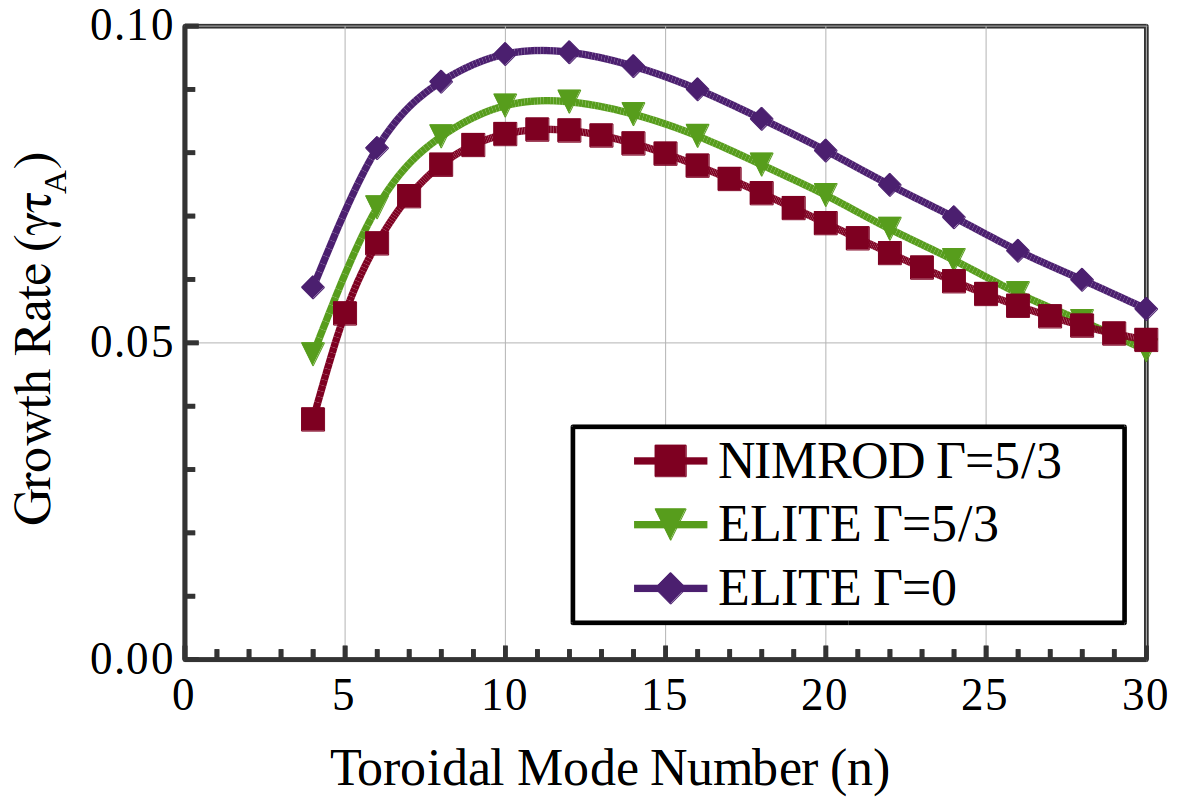}
  \vspace{-4mm}
  \caption{[Color online]
  Growth rates for the `Meudas-1' benchmark. ELITE with $\Gamma=5/3$ and
  $\Gamma=0$ are compared against results from NIMROD with $\Gamma=5/3$).
  Associated NIMROD data available in Ref.~\cite{king16Z}.}
  \label{fig:ELITEComp}
\end{figure}

\begin{figure}
  \centering
  \includegraphics[width=8cm]{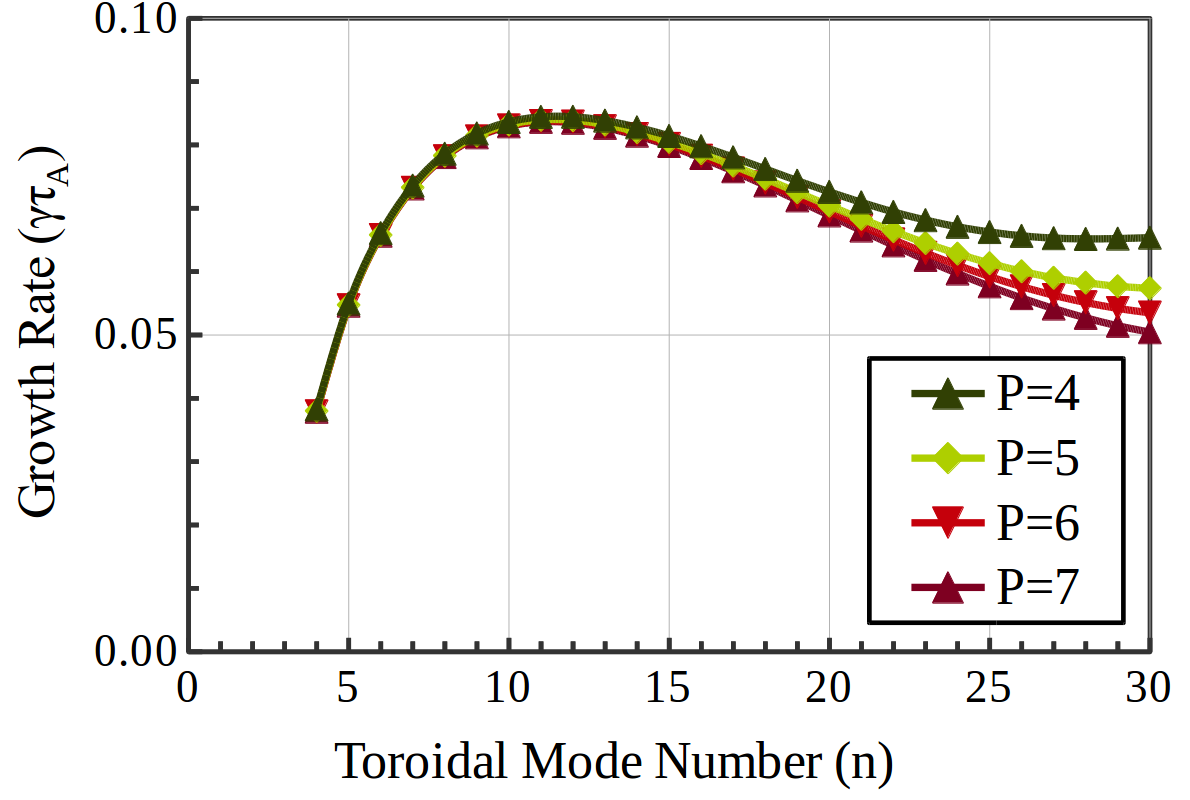}
  \vspace{-4mm}
  \caption{[Color online]
  Spectral convergence of the NIMROD code for the ideal-like parameters. 
  The maximum polynomial degree (P) of the basis functions composing the 
  spectral elements in increased in each subsequent line plotted.
  Associated NIMROD data available in Ref.~\cite{king16Z}.}
  \label{idealConv}
\end{figure}

The normalized growth rates ($\gamma \tau_A$ where the linearized mode grows as
$\text{exp}[\gamma t]$) vs.~toroidal mode number ($n_\phi$) from NIMROD and ELITE are
plotted in Fig.~\ref{fig:ELITEComp}.  There is good agreement between the
codes except for $n_\phi$=4 where there is a 27\% relative difference. All
other cases have a relative difference of less than 8\% with typical
differences of 5\%. The NIMROD convergence in terms of the maximum polynomial
order of the spectral elements is shown in Fig.~\ref{idealConv}. Convergence is
most challenging at high wavenumbers where the resolution requirements are most
stringent (the poloidal mesh is composed of $72\times512$ spectral elements).
%SEK: Great point, but total troll bait for reviewers
These cases converge from the unstable side where the growth rate decreases with
enhanced resolution. Thus the excellent agreement between NIMROD and ELITE at
high $n_\phi$ in Fig.~\ref{fig:ELITEComp} may be spurious and indicate that
slightly more resolution is required for $n_\phi$>25, however, the 
shown growth rates are likely within 5\% of their converged values.
Studying nearly ideal cases with extended MHD codes such as NIMROD is challenging 
given the vanishingly small dissipation operators, and convergence is achieved
more quickly with the additional non-ideal terms in the extended-MHD equations,
as in the cases in Sec.~\ref{sec:xMHD}.

% Relative to modeling with extended MHD, ideal-MHD convergence is more challenging 
% given the vanishingly small dissipation operators and convergence is 
% achieved more quickly with all other model equations shown in this work.

\begin{figure}
  \includegraphics[width=8cm]{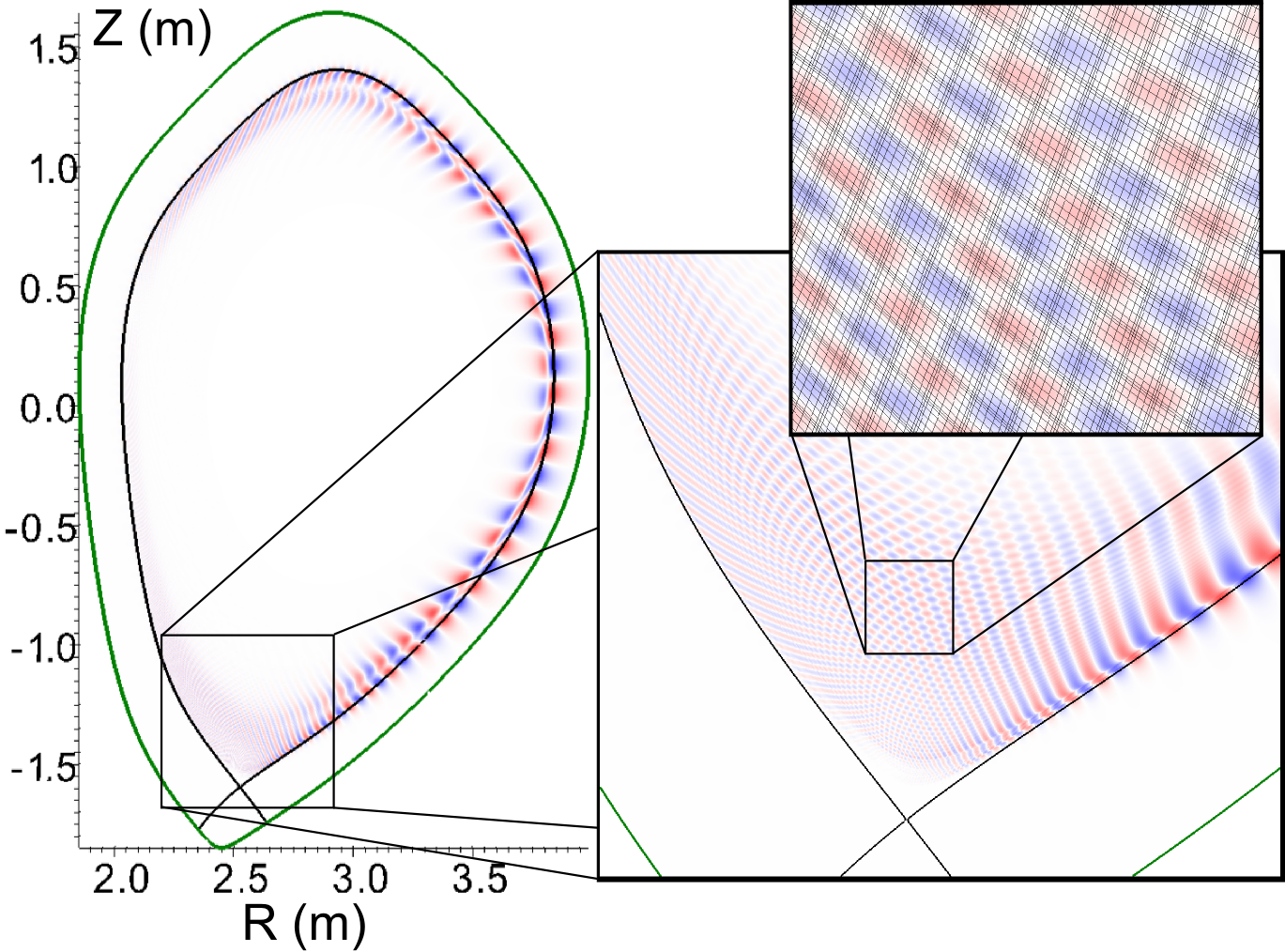}
  \caption{[Color online]
  Poloidal cross section of the radial magnetic field component of the
  $n_\phi=11$ peeling-ballooning mode from the `Meudas-1' benchmark case. }
  \vspace{-4mm}
  \label{meudas_n11_BR}
\end{figure}

Figure \ref{meudas_n11_BR} shows a poloidal cross section of the magnetic
($B_R$) eigenmode.  The mode develops an `interference-pattern' structure near
the X-point when inboard and outboard finger-like structures overlap. The
finite-element-mesh nodes are superimposed atop the smallest-scale sub-figure.
As established by Fig.~\ref{idealConv}, this simulation is spatially
and temporally converged. The high resolution required to resolve these
high-$q_{95}$, diverted cases
motivated development of memory-scaling improvements in the NIMROD code.
   
\section{Increasing the model complexity}
\label{sec:xMHD}

% o Don't emphasize M3D-C1, just ref Ferraro
% o 1fl effects:
%  - Describe profile choice for n - avoid ITG like modes w/ 2f parameters
%  - Using nu_perp = k_perp = D_n = elecd(psin=1)
%  - Using Z=3; pe_frac=0.75 => Te = Ti; Gamma=5/3
%  - Effect of density profile - local gamma * tau_a const,
%      -> normalized to core tau_a (unchanged)
%      -> local tau_a ~ sqrt(n) smaller => growth rates uniformly larger
%  - Effect of resistivity profile: high-n destabilization, low-n stabilization
% o parallel effects:
%  - Almost no effect on gamma
%  - expected from ELITE results which show mode \gamma is largely 
%      insensitive to \Gamma (0 in Ferraro, 5/3 and inf here)
% o 2fl FLR effects:
%  - Including full FLR model: Gen. Ohm's law; ion gyrovisc; cross heat fluxes
%  - Drift stabilization at high-n; crosses ideal results @ n~24
%  - Little effect at low n (n<10)
%  - (No destabilization as predicted by HRP)
%  - For this case: computations w/ and w/out x-heat flux give same result
%     -> Modest beta (0.1%); consistent with FLR theory from King and Kruger
% o Reference: Ping, Xu13

Initial-value extended-MHD computations, while more computationally expensive
than ideal-MHD calculations with ELITE, are able to include additional effects
that are more representative of experimental conditions:
% Initial-value extended-MHD computations, while more computationally expensive
% than ideal eigenvalue calculations such as ELITE, are able to include
% additional effects that are more representative of experimental conditions:
\begin{itemize} \cramplist \zapspace
\item 
Instead of a magneto-static vacuum outside the LCFS, NIMROD includes a
cold-plasma region.
\item
Finite-dissipation effects such as resistivity, viscosity and thermal
conduction.
\item 
Density and temperature profiles that prescribe the associated
drifts and resistivity profile.
\item
Parallel-closure effects that represent large diffusivities oriented along the
magnetic field.
\item
First-order FLR-closure and two-fluid effects.
\end{itemize} 
Reference \cite{Ferraro10} investigates the first three of these effects with
the Meudas-1 case. We further this work by including the last two effects.  The
first three effects are also re-investigated and different modifications to the
growth rates are found. These differences are expected as slightly different
profiles for density and temperature are used in order to avoid spurious
electrostatic modes with the two-fluid, first-order FLR model. 
Ref.~\cite{Ferraro10} avoided these modes by employing only the
single-fluid resistive-MHD model.
% {\bf SEK: I think this is review bait:  Why didn't
%   Nate see spurious modes?  What is the source of these spurious modes?
% Etc.  I don't see a clean way of discussing this}

In calculations beyond the ideal model, the density profile is
$n_e(\psi_n)=n_{e0} (p(\psi_n)/p_0)^{0.3}$ where
$n_{e0}=6\times10^{19}\;m^{-3}$ and $\psi_n$ is the normalized poloidal flux.
Here $p$ is the pressure as prescribed by the ideal gas law, $p=n_e T_e+ n_i
T_i$, where $T_\alpha$ is a species' temperature. This choice does not match
Ref.~\cite{Ferraro10}, but effectively avoids spurious electrostatic modes with
the two-fluid, first-order FLR model. These spurious modes are discussed in 
more detail in Sec.~\ref{sec:densityScan}.
% {\bf SEK: Ditto the previous comment} 
The pressure-profile gradient is specified by the Meudas-1 case and setting the
edge temperature to $200 eV$ leads to a core temperature of $5806 eV$.  The
edge temperature is modified independent of the MHD-stability (determined by
$p^\prime$) by the transformation $p(\psi_n) \rightarrow p(\psi_n) + p_c$ where
$p_c$ is a constant. Without scrape-off-layer (SOL) profiles of density and temperature, which are
not included in standard reconstructions, this choice of the edge temperature
also specifies the temperature outside the LCFS.  SOL profiles are required to
get both the correct profiles outside the LCFS, which determines the vacuum
response, and simultaneously set the correct resistivity at the mode resonant
location when Spitzer resistivity is used.  The non-ideal viscous, conductive
and particle diffusivities are set to a small value, one tenth of the
resistivity at the LCFS. 

\begin{figure}
  \centering
  \includegraphics[width=8cm]{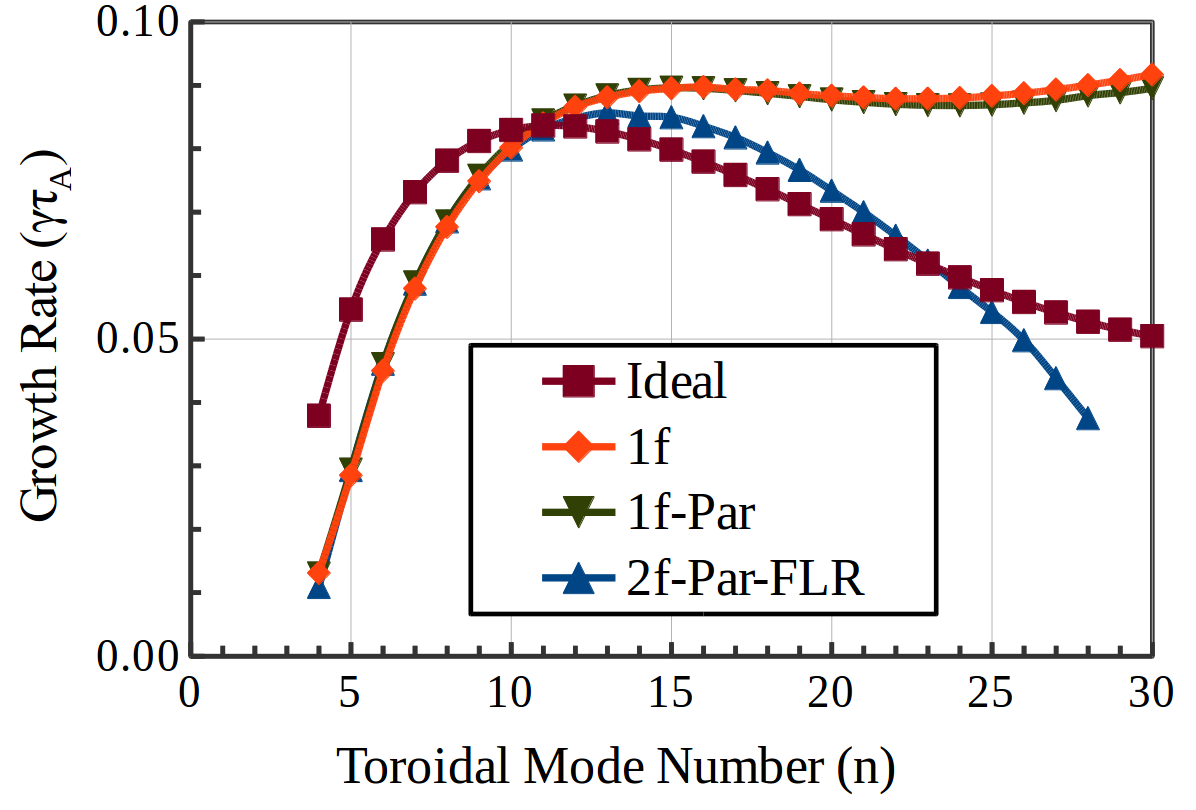}
  \vspace{-4mm}
  \caption{[Color online]
  Growth rates computed by NIMROD on the `Meudas-1' case with four different
  models: ideal MHD (ideal), resistive MHD with reconstructed density,
  temperature, and Spitzer-resistivity profiles (1f), the 1f model with parallel
  Braginskii closures (1f-Par), the 1f-Par model with ion gyroviscosity, a full
  extended MHD Ohm's law (including the Hall, $\nabla p_e$, and electron inertia
  terms) and separate temperature evolution equations with cross-heat fluxes
  (2f-Par-FLR).
  Associated NIMROD data available in Ref.~\cite{king16Z}.}
  \label{modelScan}
\end{figure}

Figure~\ref{modelScan} plots the growth rates computed by NIMROD on the
`Meudas-1' case with four different models: ideal MHD (ideal), resistive MHD with
reconstructed density, temperature, and Spitzer-resistivity profiles (1f), the
single-fluid model with parallel Braginskii closures (1f-Par) where the 
parallel viscosity is
\begin{multline}
\mathbf{\Pi}_{\parallel i}=
  m_i n_i \nu_{\parallel i}
  \left(\hat{\mathbf{b}}
  \hat{\mathbf{b}}-\frac{1}{3}\mathbf{I}\right) \\
  \times\left(3\hat{\mathbf{b}}\cdot\nabla\mathbf{v}_{i}\cdot
  \hat{\mathbf{b}}-\nabla\cdot\mathbf{v}_{i}\right)\;,
\end{multline}
and the parallel heat flux vector is
\begin{equation}
\mathbf{q}_{\parallel \alpha}=
  -n_i \chi_{\parallel \alpha}
  \hat{\mathbf{b}}\hat{\mathbf{b}}\cdot\nabla T_{\alpha}\;,
\end{equation}
and a two-fluid model with parallel closures that includes ion gyroviscosity
\begin{multline}
\mathbf{\Pi}_{\times i}=
  \frac{m_{i}p_{i}}{4ZeB}[\hat{\mathbf{b}}
    \times\mathbf{W}\cdot
    \left(\mathbf{I}+3\hat{\mathbf{b}}\hat{\mathbf{b}}\right) \\
    -\left(\mathbf{I}+3\hat{\mathbf{b}}\hat{\mathbf{b}}\right)
    \cdot\mathbf{W}\times\hat{\mathbf{b}}]\;,
\end{multline}
where $\mathbf{W}$ is the rate-of-strain tensor
\begin{equation}
\mathbf{W}=\nabla\mathbf{v}+\nabla\mathbf{v}^{T}
           -(2/3)\mathbf{I}\nabla\cdot\mathbf{v}\;,
\end{equation}
a full extended-MHD Ohm's law,
\begin{equation}
\mathbf{E}=
  -\mathbf{v}\times\mathbf{B}+\frac{\mathbf{J}\times\mathbf{B}}{n_ee}
  -\frac{\nabla p_{e}}{n_ee}+\eta\mathbf{J}
  +\frac{m_{e}}{n_e e^2}\frac{\partial\mathbf{J}}{\partial t}\;,
\end{equation}
and separate temperature evolutions with cross-heat fluxes,
\begin{equation}
\mathbf{q}_{\times \alpha}=
  \frac{5p_{\alpha}}{2q_{\alpha}B}
  \hat{\mathbf{b}}\times\nabla T_{\alpha}\;,
\end{equation}
(2f-Par-FLR). Here $v_i$ is the bulk-ion flow, $\nu_{\parallel i}$ is the ion parallel
diffusivity, $\chi_{\parallel \alpha}$ is the species' parallel diffusivity,
$q_\alpha$ ($e$) is the species' (electron) electric charge,  $\mathbf{J}$ is
the current density, $\hat{\mathbf{b}}$ is the magnetic unit vector
($\hat{\mathbf{b}} = \mathbf{B}/B$), and $\mathbf{I}$ is the identity tensor.
An effective ion charge ($Z$) of three is assumed. Each model builds
on the previously listed and includes all prior terms. Electron viscosity is
not included.

% REMOVED FROM INTRO
% For peeling-ballooning modes, the prevailing analytic description of these
% effects is described in Ref.~\cite{Hastie03}. 
% This conventional picture expects two-fluid and FLR effects are destabilizing
% to modes with low toroidal mode number but stabilizing at high toroidal mode
% number. 
% We compare our results with this simple model of drift stabilization where
% $\gamma_{MHD}^2 = \omega (\omega - \omega_*)$.  Here $\gamma_{MHD}$ is the
% ideal-MHD growth rate, $\omega$ is the complex mode frequency and $\omega_*$ is
% the diamagnetic-drift frequency.

% Closure expressions for the viscosity tensor, $\mathbf{\Pi}$, and heat flux,
% $\mathbf{q}$. These expressions are typically decomposed into three parts:
% parallel (large diffusivities oriented along the magnetic field), cross (FLR
% ordered contributions), and perpedicular (small dissipative terms)

When a density profile is included, the normalized growth rate of the mode
remains constant with the single-fluid model when the Alfvén time is computed
at the mode resonant surface. Our calculations are normalized by the Alfvén
time as computed with the values at the magnetic axis. These values remain constant
when a density profile is included such that the normalized mode growth rate is
effectively increased. 

The effect of the resistivity profile is more nuanced.  Finite
resistivity allows for reconnection within the model by relaxing the
frozen-flux constraint. This permits resistive-ballooning modes that
grow faster than their ideal counterparts. Alternatively, the resistive
dissipation can stabilize the modes by acting as a dissipative term and
modifying the response outside the LCFS from that of a vacuum to that of
a plasma. Figure 5 of Ref.~\cite{Ferraro10} shows that the plasma
response is stabilizing relative to a vacuum region. Overall, these
effects stabilize the ballooning modes at low-$n_\phi$ and destabilize
the modes at high-$n_\phi$ as seen in Fig.~\ref{modelScan} when
comparing the single fluid and ideal normalized growth rates. The effect
of including small particle, viscous, and thermal diffusivities on the
resistive-MHD mode is small (not shown).

Including the Braginskii parallel closures with large coefficients
($\chi_{\parallel e} = 10^9\times \eta_0/\mu_0$ and  $\chi_{\parallel i} =
\nu_{\parallel i} = 10^8\times \eta_0/\mu_0$) has little effect on the growth
rates.  When thermal conduction is large, the temperature quickly equilibrates
along field lines to produce an isothermal response.  Thus NIMROD computations
that show little effect from the parallel closures are consistent with the ELITE
results that find the shape of the mode growth-rate spectrum is largely not
sensitive to the value of the ratio of specific heats, $\Gamma$, as seen in
Fig.~\ref{fig:ELITEComp}. When the response to the compressible motion
associated with the sound wave is eliminated ($\Gamma=0$ in the ELITE 
computations), the growth rate is slightly enhanced relative to the adiabatic limit.
If the growth rates are modified beyond this small effect, it would be
an indication that changes in the energy equation can impact the MHD response, 
and thus that closure effects are significant. Although the short-mean-free
path Braginskii-like closure \cite{Braginskii,Catto04} is used in our NIMROD
compuatations, we note that other closures, such as long-mean-free path
\cite{Ramos11} or the general approach of solving drift-kinetic equation
\cite{held15} for a closure can be applied. However, for this case where the
parallel-closure effects are negligible, different parallel closures are
unlikely to significantly modify the results.

With the full two-fluid, FLR model, there is a stabilizing effect on the
intermediate and high-$n_\phi$ modes as expected from analytic
treatments \cite{Hastie03}. A common reduced-MHD approximation for
first-order FLR closures is to assume that the ion-gyroviscous force and
the advection by the diamagnetic drift exactly cancel (this is
colloquially known as the gyroviscous cancellation, see, e.g.
\cite{Coppi64}). However, as discussed by Ramos in
Ref.~\cite{Ramos:2007cn}, ``these cancellations are only partial and not
very useful in practice for general magnetic geometries\ldots''.  This
cancellation is only valid with a large, uniform guide field, without
curvature terms (slab approximation), and in the low-$\beta$ limit (see,
e.g. \cite{King14}).  Use of the full gyroviscous operator is critical
as in addition to effects that appear as diamagnetic drifts, it also
contains terms proportional to first-order FLR magnetic-curvature and
grad-B drifts \cite{King11}. Our modeling contains not just the
first-order FLR drifts from ion gyroviscosity, but also the drifts from
the cross-heat fluxes that enter the equations on the same order.  

Just as the large balance of the ideal-MHD forces can cause
numerical difficulties with the ideal-MHD force operator, the
partial gyroviscous cancellation leads to similar numerical 
pitfalls and verification is important.
The implementation of these terms in NIMROD are verified in cylindrical and
slab magnetic geometries against analytic calculations for
tearing~\cite{Sovinec10,King11,King14} and
ion-temperature-gradient~\cite{Schnack13} (ITG) modes. 
% {\bf SEK: I don't know
%   why you have Schnack13 associated with tearing instead of ITG.  Also,
%   Schnack really didn't use Coppi67 directly but rather had to modify it
%   for the verification purposes, but it's better to just reference his
% paper}.  
% JRK -> previous references Coppi67 for ITG; others for general verfication
To test the effect of these terms against analytic theory
in full tokamak magnetic geometry, we perform a qualitative
comparison with analytic theory in Sec.~\ref{sec:analyticComp}.
Full verification is precluded by the approximations
made in the analytic theories.
%For verification in full tokamak magnetic geometry, we perform a qualitative
%comparison with analytic theory in Sec.~\ref{sec:analyticComp}.

The first-order FLR model is valid for the toroidal modes presented although
this is not necessarily the case for general edge mode modeling.  Assuming
$k\simeq(q/r+1/R)n_\phi$ and evaluating local quantities at the peak of the
mode eigenfunction ($\psi_n = 0.969$) on the outboard midplane, the normalized
ion-gyroradius ($\rho_i=\sqrt{\Gamma m_i T_i}/ZeB$) as a function of toroidal
mode is $k \rho_i \simeq 0.0055n_\phi$.  If the first-order FLR assumption is
violated, more-complex full-ion-orbit kinetic modeling is required to simulate
large-fluctuation ELM dynamics.  For this case, computations produce
approximately the same results with and without the cross-heat fluxes.  This is
consistent with drift analytics \cite{King14} that shows that the cross-heat
flux is only significant at very low values of plasma $\beta$ (here the
$\beta=2\mu_0p/B^2$ is 0.1\%).  Perhaps coincidentally for this case, the
effects of the resistive destabilization and drift stabilization largely
counteract one another. The drift stabilized growth rates are not less than the
ideal calculation until $n_\phi \gtrsim 24$.  

% Old text:
%
% In addition to the ideal MHD benchmark, the M3D-C1 code
% investigated the effects of including a density gradient profile and Spitzer
% resistivity. Our NIMROD studies have furthered this work to
% include two-fluid effects through the generalized Ohm's law (including the
% Hall, $\nabla p_e$, and electron inertia terms), ion gyroviscosity, highly
% anisotropic thermal conduction, and parallel ion
% viscosity.  When the model includes a generalized Ohm's
% law, the result qualitatively agrees with the analytic theory of
% Ref.~\cite{Hastie03} used for ELITE's {\em ad hoc} model; we find a
% destabilization of the intermediate-$n$ mode spectrum and a stabilization of
% the high-$n$ mode spectrum. When ion gyroviscosity is included, there is a
% stabilization of the high-$n$ mode spectrum, consistent with recent results
% from the BOUT++ code \cite{Xu13}. The high-$n$ mode
% spectrum that results from the use of a extended MHD model prevents energy from
% accumulating on the smallest resolvable scales and corrupting the simulations.
% These results are shown in Fig.~\ref{fig:ELITEComp}.

\section{Comparison to drift analytics}
\label{sec:analyticComp}

\begin{figure}
  \centering
  \includegraphics[width=8cm]{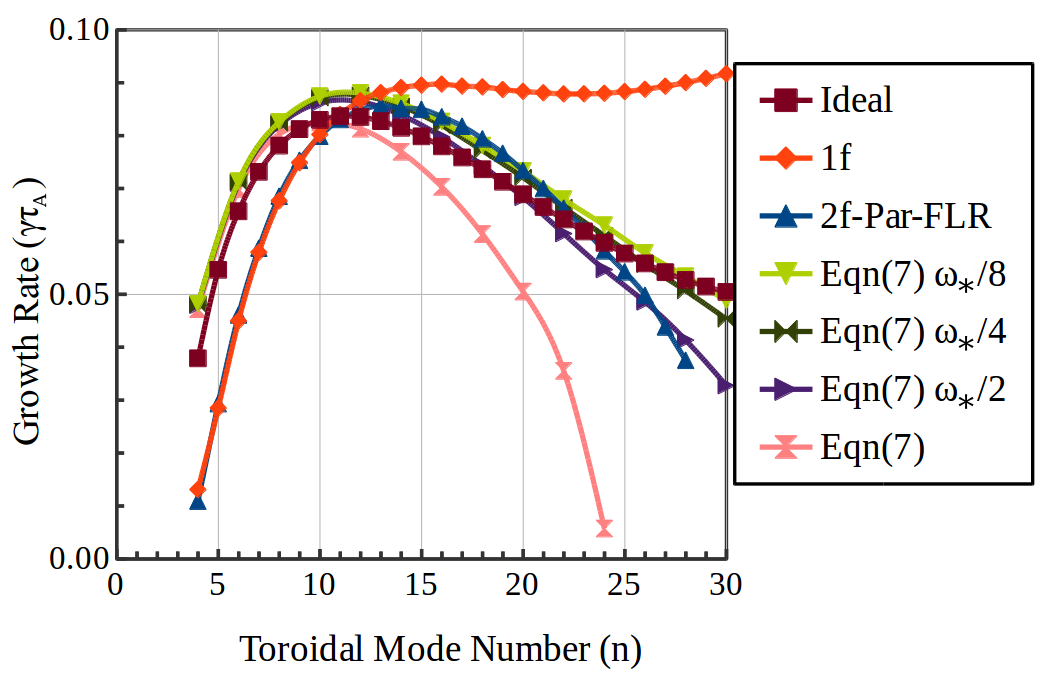}
  \vspace{-4mm}
  \caption{[Color online]
  Growth rates computed by NIMROD with the ideal, single-fluid (1f) and full
  FLR (2f-Par-FLR) models in comparison to the analytic dispersion relation of
  the FLR-stabilized, ideal ballooning mode with full and reduced $\omega_{*\alpha}$ 
  values. Associated NIMROD data available in Ref.~\cite{king16Z}.}
  \label{FLRidealComp}
\end{figure}

To gain insight into our two-fluid computations, we compare with analytic
descriptions of drift stabilization of the ballooning mode.  The dispersion
relation for drift stabilizations of the ideal, incompressible ballooning mode
is
\begin{equation}
\omega (\omega - \omega_{*i}) = - \gamma_I^2
\label{eq:driftBalloon}
\end{equation}
where $\gamma_I$ is the ideal growth rate in the absence of drift effects and
$\omega$ is the complex frequency ($\omega=i\gamma$) \cite{tang82}.  The
diamagnetic drift velocity is typically defined as the
perpendicular-to-the-magnetic-field flow contribution from the pressure
gradient within the context of a two-fluid, or generalized Ohm's law.  For a
tokamak, neoclassical effects damp the poloidal diamagnetic flow contribution.
Thus in our case the remaining toroidal diamagnetic flow determines the
diamagnetic frequency, $\omega_{*i}= \mathbf{k}\cdot\mathbf{v}_{*i} =
n_\phi/(n_iq_i)\partial p_i/\partial \psi$ where $\mathbf{k}$ is the mode
wavenumber.  Figure \ref{FLRidealComp} shows the comparison to this dispersion
relation with full and reduced diamagnetic drift values. The value of
$\omega_{*i}$ is computed at the radial location where the mode structure
peaks.  The normalized frequency from the diamagnetic drift velocity,
$\omega_{*i}\tau_A$, varies linearly with toroidal mode number as
$0.00483n_\phi$.  This `local' approximation, which neglects the radial
variation of the $\omega_{*i}$ profile, is known to over-estimate the effect of
drift stabilization \cite{Hastie00,Snyder11}.  We find that a reduction of the
value of $\omega_{*i}$ (approximately by a factor of 2-4) gives the best
agreement between Eqn.~\eqref{eq:driftBalloon} and the two-fluid NIMROD
calculation.  A similar comparison of ELITE to the two-fluid BOUT++ model on a
different case finds the drift stabilization is over-predicted \cite{Snyder11}
by a comparable factor.

One of the limitations of Eqn.~\eqref{eq:driftBalloon} is that it does not
include the effect of resistive destabilization on the high-$n_\phi$ modes,
as is included in our two-fluid computations.
The analytic treatment of Ref.~\cite{Hastie03} provides a dispersion relation
(Eqn.~(43) of the reference; referred to as HRP Eqn.~(43) in this discussion) that
includes the effect of resistive destabilization in addition to drift effects.
In order to compare with this theoretical treatment, parameters are computed
on the outboard midplane at the radial location where the mode structure peaks,
consistent with the previous $\omega_{*i}$ calculation. This yields local values
of the safety factor, $q=7.34$, normalized magnetic shear, $s=8.8$ and
normalized pressure gradient, $\alpha=28.8$, where the definitions of
Ref.~\cite{Hastie03} are used.  The $\Delta^\prime_B$ stability parameter is
determined by using the ideal growth rate ($\Gamma=5/3$) and solving Eqn.~35 of
Ref.~\cite{Hastie03}.  $\Delta^\prime_B$ varies with toroidal mode number but
falls in the range of $-17.4$ to $-12.1$. 

\begin{figure}
  \centering
  \includegraphics[width=8cm]{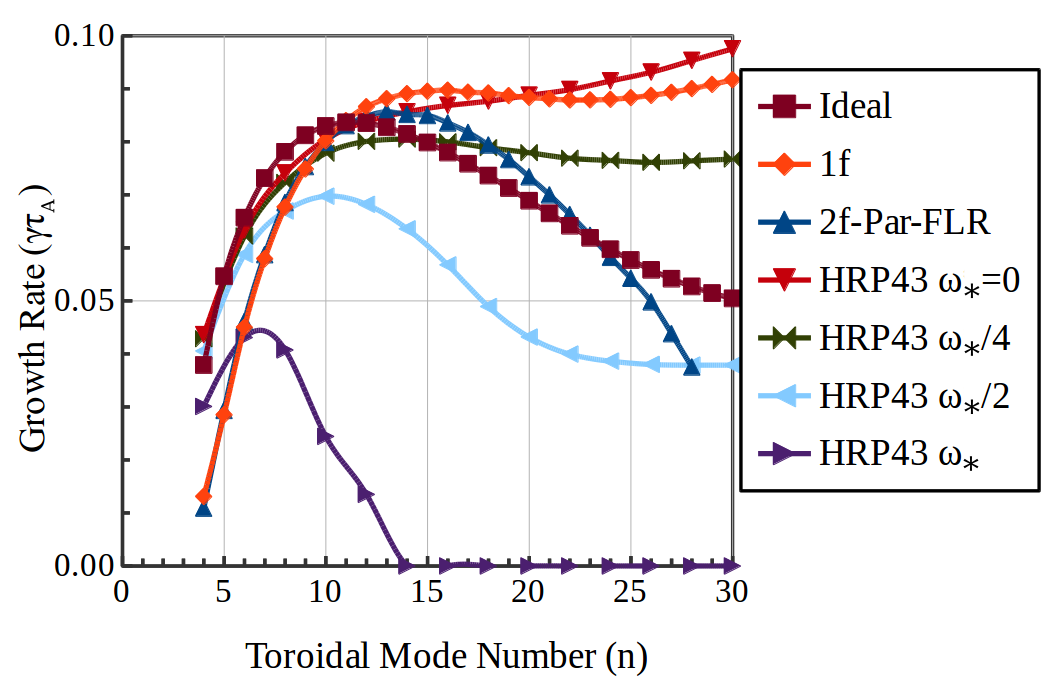}
  \vspace{-4mm}
  \caption{[Color online]
  Growth rates computed by NIMROD with the ideal, single-fluid (1f) and full
  FLR (2f-Par-FLR) models in comparison to the analytic dispersion relation of
  Ref.~\cite{Hastie03}, Eqn.~(43) with the same parameters as the NIMROD
  computations except reduced $\omega_{*\alpha}$ values and a reduced
  resistivity (0.35 of the NIMROD value).
  Associated NIMROD data available in Ref.~\cite{king16Z}.}
  \label{HRPcomp}
\end{figure}

Comparison of HRP Eqn.~(43) with $\omega_*=0$ and the NIMROD resistive-MHD results
differ by a factor approximately 2.5 at large toroidal mode numbers where the
analytic calculation results in greater growth rates and there is no
stabilization at low $n_\phi$. As the analytic calculation is a local calculation, it
does not capture the stabilizing influence on the global mode of finite
resistivity, in particular it misses the effect of modifying the response outside the
LCFS from that of a vacuum to that of a resistive plasma.  Our interest is
largely to compare and contrast the predicted drift stabilization and thus we
reduce the resistivity to 35\% of the value used in the NIMROD computations
when calculating the analytic values of HRP Eqn.~(43). This produces growth
rates with $\omega_*=0$ that roughly match the NIMROD resistive-MHD
computations at high $n_\phi$ as shown in Fig.~\ref{HRPcomp}.

More significantly, Fig.~\ref{HRPcomp} compares the results of our linear
NIMROD computations with the drift-stabilized growth rates computed from
Eqn.~(43) of Ref.~\cite{Hastie03}. Again, the analytics of HRP Eqn.~(43)
over-estimate the effect of drift stabilization where the reduction of
$\omega_*$ that leads HRP Eqn.~(43) to qualitatively agree with the two-fluid
NIMROD computations is between 2 to 4 (dependent on $n_\phi$). There is little
effect on the low-$n_\phi$ modes and the ion drift-wave resonance predicted in
Ref.~\cite{Hastie03} is not observed.

\section{FLR parameter scan}
\label{sec:densityScan}

In order to ascertain the role of two-fluid effects further, we vary the
density and temperature while scaling parameters to maintain
constant $\beta$ and $S$ (other dissipation parameters are scaled relative to
resistivity). In experimental discharges, the zeroth-order effect on the edge
of modifying the density is modify the plasma collisionality and thus the
bootstrap current. This causes a transition from ballooning-like modes at high
density to peeling-like modes at low density. Our computations use a fixed
equilibrium current and thus are not sensitive to this effect. Instead, we
isolate the effect of the modification of $\rho_i$ (and the associated ion
skin depth, $d_i$) within the context of our two-fluid, first-order FLR model.
This contrasts with Ref.~\cite{xu14} that examines these effects in concert
and Ref.~\cite{Zhu12} that considers only the effect of modifications
to the current profile.

\begin{figure}
  \centering
  \includegraphics[width=8cm]{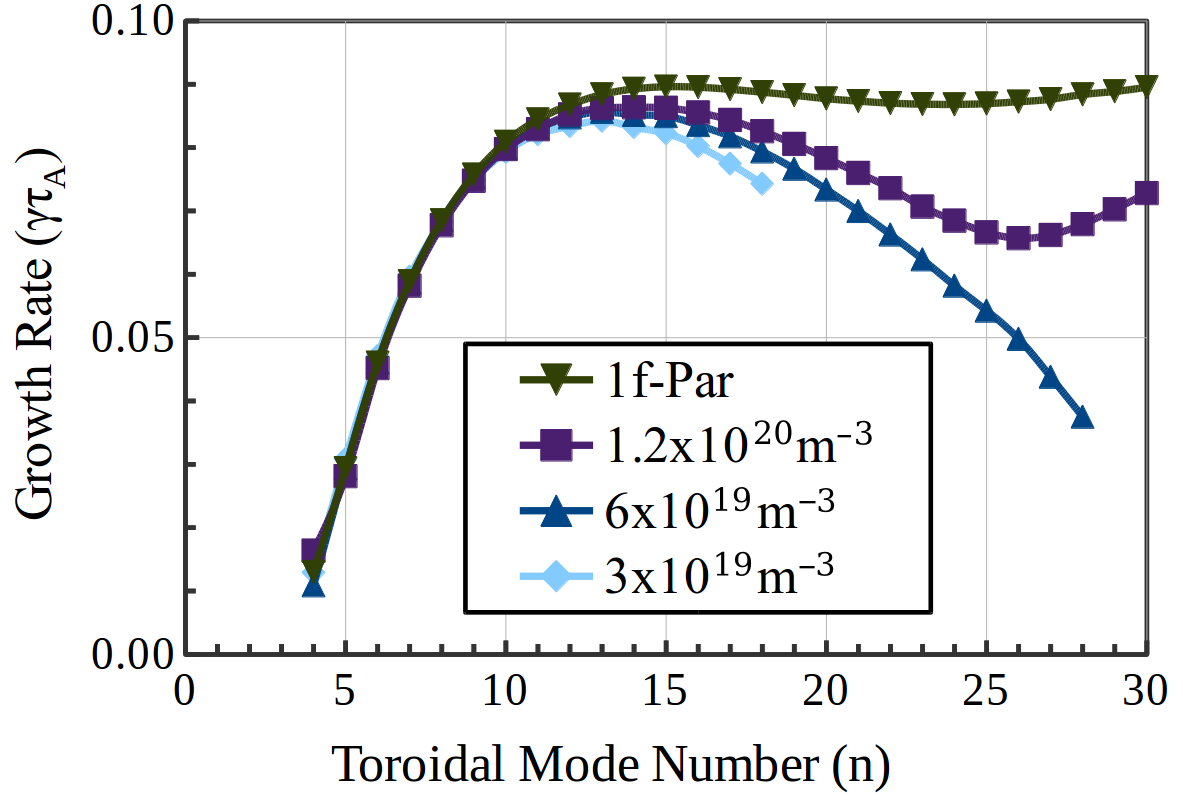}
  \vspace{-4mm}
  \caption{[Color online]
  Growth rates computed by NIMROD on the `Meudas-1' case with the 1f model
  with parallel Braginskii closures (1f-Par; single-fluid limit), and three
  different densities with the two-fluid, FLR model with parallel Braginskii
  closures (2f-Par-FLR): the core densities are $n_e=3\times10^{19}$,
  $6\times10^{19}$ and $1.2\times10^{20}\;m^{-3}$.
  Associated NIMROD data available in Ref.~\cite{king16Z}.}
  \label{densityScan}
\end{figure}

Figure \ref{densityScan} shows the linear growth rate from NIMROD computations
for resistive-MHD and three two-fluid cases with varying densities (the
core density is varied from $3\times10^{19}\;m^{-3}$ to
$1.2\times10^{20}\;m^{-3}$ where $6\times10^{19}\;m^{-3}$ is the value used in
the prior computations discussed in the paper). The effect of
drift-stabilization is stronger at low densities ($\rho_i$ scales as
$1/\sqrt{n_i}$ at constant $\beta$) as expected. The cases with
$n_e=3\times10^{19}\;m^{-3}$ are not plotted for $n_\phi>18$ as the most unstable
mode is no longer the peeling-ballooning mode but rather a dominatly
electro-static mode related to the ITG as studied in the context NIMROD
two-fluid advance in Ref.~\cite{Schnack13}. This region is excluded as
examination of ITG mode dynamics is outside the scope of this work.

% In order to ascertain the role of two-fluid effects further, we vary one of the
% two-fluid FLR parameters that set the ion gyroradius $\rho_i$: the ion skin
% depth, $d_i$, and the plasma $\beta$.  Keeping the Meudas-1 current and pressure
% profiles fixed, we scale the density and temperature profiles in such a way
% that the plasma pressure remains the same in all simulations. This modifies the
% ion skin depth at constant $\beta$. The current (including the bootstrap
% contribution) and the Lundquist and Prandtl numbers are fixed to isolate
% effects from the two-fluid, FLR model only. As shown in
% Fig.~\ref{densityScan}, the low density cases (large $d_i$, $\rho_i$) peak at
% lower $n$ and are stabilized. However, there is a high-$n$ destabilization that
% is related to the ITG drive as captured by the fluid model. This ITG drive has
% been studied further in Ref.~\cite{Schnack13}.  It is our experience that for
% reconstructed cases that include density and temperature measurements, the ITG
% drive is small - likely a result of ITG modification of the profiles to ensure
% at least marginal stability.

\section{Discussion and Conclusions} 
\label{sec:conclusions}

Our comparisons to the ideal PBM growth rates computed by the ELITE code and
analytic descriptions of the resistive, two-fluid PBM provide another
verification test of the NIMROD code as a means to study the tokamak edge.
Quantitative agreement with ELITE is achieved for these cases, extending prior
work \cite{Burke10} that did not include diverted magnetic topology.

Two-fluid and FLR-drift effects can both produce stabilizing (drift
stabilization through the sheared motion of the ion and electron fluid
responses) and destabilizing (species decoupling and enhanced growth
through mediation by the more mobile electron fluid) effects (see, e.g.,
\cite{King11,King14}).  At intermediate and large mode numbers, the
inclusion of finite-resistivity is destabilizing while the FLR-drift
effects are stabilizing. These effects are largely offsetting, resulting
in a toroidal-mode growth-rate spectrum with the full extended-MHD model
that resembles the ideal calculation. This result is not general but
rather case dependent.  Consistent with ELITE calculations that vary the
ratio of specific heats from the adiabatic through the isothermal limit, we
find the addition of anisotropic thermal conduction produces only a
small modification of the growth rate.

Relative to the NIMROD calculations, analytics \cite{Hastie03} over-predict
both the effects of resistive destabilization and FLR-drift stabilization for
the case studied. Qualitative agreement with the analytic descriptions of the
peeling-ballooning mode is found where high-$n_\phi$ modes are susceptible to
resistive destabilization and drift stabilization in both the analytics and
NIMROD computations.  One of the limits of the analytic derivation is a `local'
approximation which leads to the over prediction of FLR-drift
stabilization~\cite{Hastie00}.  Given the highly complex nature of both the
configuration of the tokamak edge and the two-fluid model equations,
quantitative agreement is not expected as the local approximation neglects the
global configuration and profile effects.

Regarding our investigation into the impact of the size of the ion gyroradius within the
context of NIMROD's full first-order FLR, two-fluid model, we find significant
changes on the growth rates of high-$n_\phi$ modes at varying densities but
constant $\beta$. However, the most unstable mode number is largely unchanged.
Thus we conclude the zeroth-order effect of modifications to the bootstrap
current largely dominates the determination of the most unstable mode during
collisionality scans while lower densities (large $\rho_i$) increase the
magnitude of the drift stabilization on the high-$n_\phi$ modes consistent
with the discussion of Ref.~\cite{Snyder07}.

\appendix
%--------------------------------------------------

\begin{acknowledgments}
We thank Carl Sovinec, Chris Hegna and Nate Ferraro for discussions involving
this paper and N.~Aiba for providing the Meudas-1 equilibrium. This material is
based on work supported by US Department of Energy, Office of Science, Office
of Fusion Energy Sciences under award numbers DE-FC02-06ER54875 and
DE-FC02-08ER54972. This research used resources of the National Energy Research
Scientific Computing Center, a DOE Office of Science User Facility supported by
the Office of Science of the U.S. Department of Energy under Contract
No.~DE-AC02-05CH11231.
\end{acknowledgments}
\bibliographystyle{apsrev4-1}
\bibliography{Biblio}

\end{document}